\documentclass{webofc}
\usepackage[varg]{txfonts}   % Web of Conferences font
\usepackage{graphics}
\usepackage{graphicx}
\usepackage{subfig}
\usepackage{wrapfig}

\usepackage{amsmath}
\usepackage{amsfonts}
\usepackage{amssymb}

\graphicspath{ 
  {Images/} 
}

\begin{document}
\title{Open and hidden heavy flavor measurements at RHIC}

\author{\firstname{Leszek} \lastname{Kosarzewski}\inst{1}\fnsep\thanks{\email{kosarles@fjfi.cvut.cz}}}

\institute{Faculty of Nuclear Sciences and Physical Engineering, Czech Technical University in Prague, Břehová 78/7, 115 19 Staré Město, Prague, Czech Republic}

\abstract{%  
Quarks of heavy flavors are useful tool to study quark-gluon plasma created in heavy-ion collisions.
Due to their high mass and early production time, heavy quarks experience the entire evolution of the system created in these collisions.
Open heavy flavor meson measurements are sensitive to the energy loss in the QGP, while quarkonia are sensitive to the temperature of the QGP as they dissociate because of Debye-like screening of color charges.

This presentation is a summary of the latest heavy flavor studies performed at RHIC. 
Results from both STAR and PHENIX experiments are presented, compared to theoretical calculations and the implications discussed.
%Both STAR and PHENIX experiments had programs of heavy flavor measurements with their dedicated precise vertex detectors. These results of these studies will also be discussed.
}

\maketitle
\section{Introduction}
\label{intro}

Heavy quarks are a good probe of quark gluon plasma properties, which can be created in relativistic heavy ion collisions, thanks to their early production and treatment with pQCD calculations. The production mechanisms and in-medium interactions are however different for open heavy flavor mesons and quarkonia. The first group may be produced via fragmentation or coalescence, while interactions with the hot medium include energy loss due to induced gluon radiation~\cite{GYULASSY1990432} or elastic collisions~\cite{THOMA1991491}. The radiative energy loss is mass dependent and resulted in "`dead cone"' effect which was directly observed by ALICE~\cite{DeadConeEffect_obs}. Quarkonia, on the other hand, are produced through color singlet~\cite{bib:Onium:CS:1980} or color octet~\cite{bib:Onium:CO:Bmeson} intermediate states which hadronize into a bound state of $q\bar{q}$. The color octet channel requires color neutralization by emission of additional gluons. In QGP, quarkonia are affected by screening of color charges, which causes the bound state to dissociate at high temperature~\cite{matsui}. Furthermore quarkonium states can also be supressed due to parton energy loss in the medium~\cite{bib:Onium:Eloss}. If the density of heavy quarks is sufficiently high, it is also possible for quarkonium to regenerate~\cite{bib:Ups:Rapp}.

Measurements of heavy flavor spectra in $p+p$ collisions allow to test the production models. In order to study the suppression within the QGP, a nuclear modification factor $R_{AA}$ has to be measured. Elliptic flow is studied with $v_{2}$ coefficient and provides information about the interactions with the medium and the degree of thermalization. Finally, studies of $p+A$ or $d+A$ collisions allow to test the influence of cold nuclear matter effects on the heavy quark production~\cite{bib:Jpsi:Shadowing}.

This paper focuses on recent heavy flavor measurements performed at RHIC by STAR and PHENIX experiments.

\section{Charm and bottom production}

\begin{figure}[h]
\centering
	\subfloat{		
		{\includegraphics[width=0.46\textwidth]{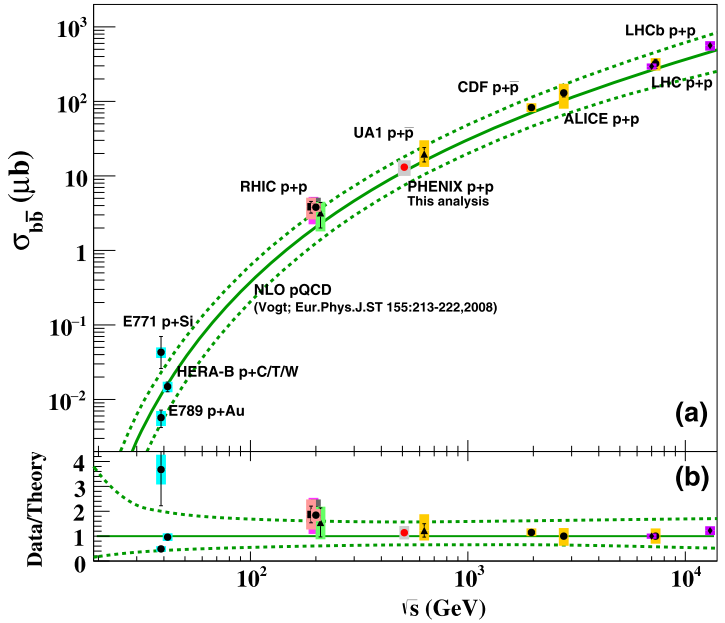}
		\label{fig:ccbb:PHENIX:bb}}
		{\includegraphics[width=0.54\textwidth]{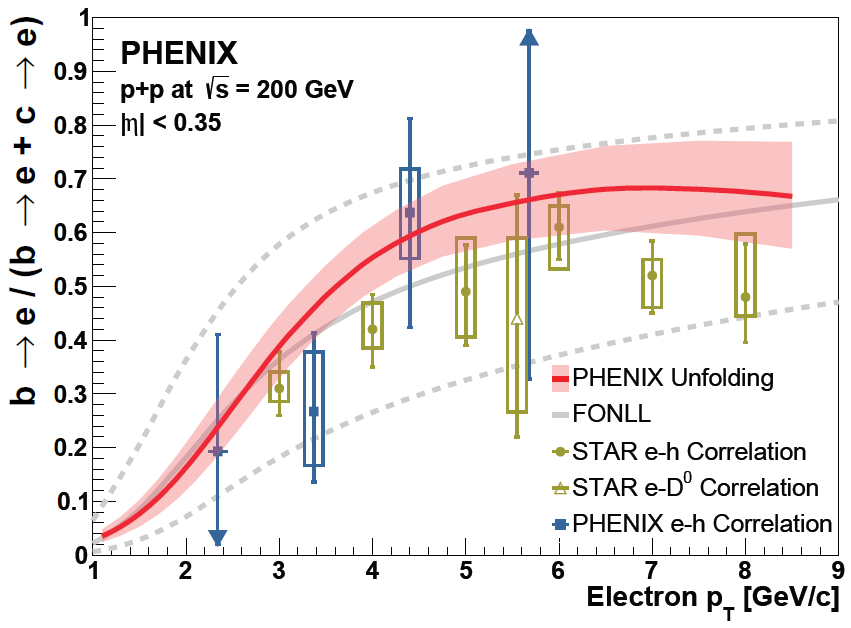}
		\label{fig:ccbb:PHENIX:bfrac}}
		}
\caption{Left: Bottom production cross section vs. $\sqrt{s}$~\cite{bib:bb_oscil:PHENIX:pp500} compared to NLO pQCD calculation~\cite{bib:Vogt_2008:ccbb}.
Right: Fraction of semileptonic decays of $b$ to the total $c+b$ semileptonic decays~\cite{bib:cbe:PHENIX:pp}.}
\label{fig:ccbb}       % Give a unique label
\end{figure}

Production of charm and bottom in $p+p$ collisions was measured both by STAR~\cite{bib:STAR:Drun09} and PHENIX~\cite{bib:cbe:PHENIX:pp,bib:cbe:PHENIX:dAu,bib:cbe:PHENIX:ppAuAu} experiments, which allowed the calculation of charm and bottom production cross sections. Especially interesting is a measurement of $b\bar{b}$ cross section via $B^{0}\leftrightarrow\bar{B^{0}}$ meson oscillations~\cite{bib:bb_oscil:PHENIX:pp500}. This is done by selecting like-sign dimuons originating from the aforementioned oscillations. The measured spectrum favors $b\bar{b}$ production through flavor creation or excitation over gluon splitting. The overall trend of production cross section vs. collision energy is well described by NLO pQCD calculations~\cite{bib:Vogt_2008:ccbb} as shown in Fig.~\ref{fig:ccbb:PHENIX:bb} (left).

\subsection{Charm hadrochemistry}

\begin{figure}[h]
\centering
	\subfloat{		
		{\includegraphics[width=0.53\textwidth]{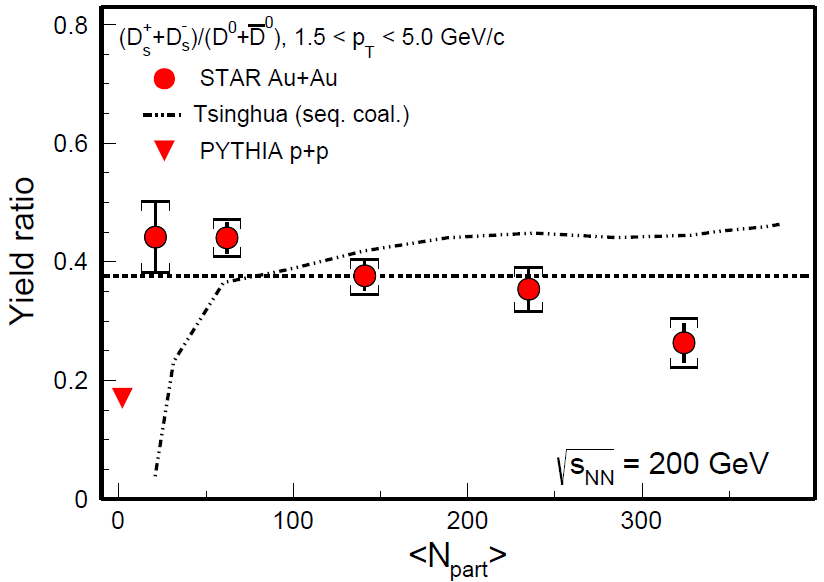}
		\label{fig:charm:STAR:chemistry:D}}
		{\includegraphics[width=0.47\textwidth]{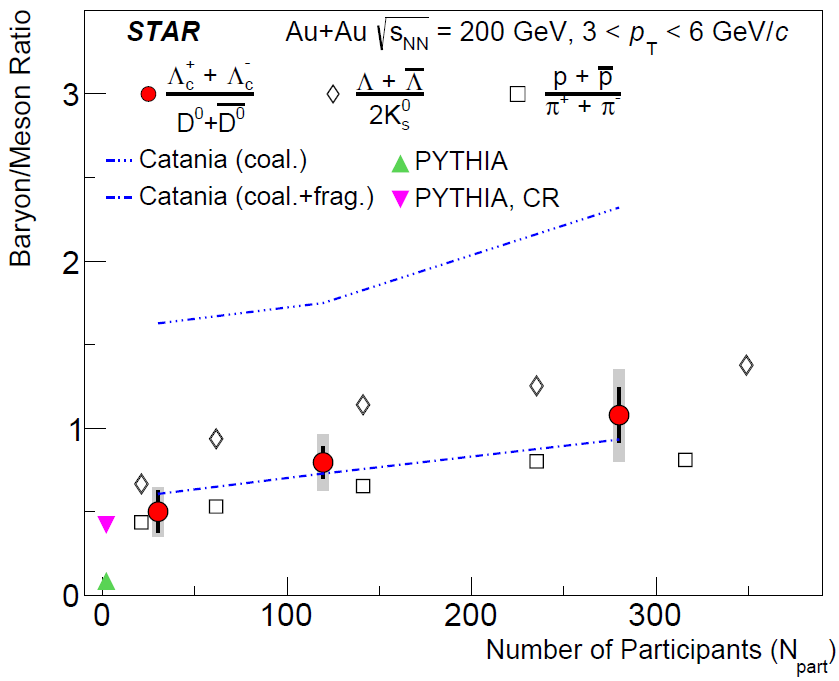}
		\label{fig:charm:STAR:chemistry:lambdac}}
		}
\caption{Left: Ratio of $(D^{+}_{s}+D^{-}_{s})/(D^{0}+\bar{D}^{0})$~\cite{bib:STAR:DsToD0} vs. $<N_{part}>$ measured by STAR at $200\:\mathrm{GeV}$.
Right: Ratio of $(\Lambda^{+}_{c}+\Lambda^{-}_{c})/(D^{0}+\bar{D}^{0})$ vs. $<N_{part}>$~\cite{bib:STAR:lambda_c}.
}
\label{fig:charm:STAR:chemistry}       % Give a unique label
\end{figure}

Study of charm hadrochemistry in $Au+Au$ collisions allows to investigate the hadronization mechanisms of these quarks. STAR measured ratios of $(D^{+}_{s}+D^{-}_{s})/(D^{0}+\bar{D}^{0})$~\cite{bib:STAR:DsToD0}, shown in Fig.~\ref{fig:charm:STAR:chemistry:D} (left), and $(\Lambda^{+}_{c}+\Lambda^{-}_{c})/(D^{0}+\bar{D}^{0})$ at $200\:\mathrm{GeV}$~\cite{bib:STAR:lambda_c}, presented in Fig.~\ref{fig:charm:STAR:chemistry:lambdac} (left) as a function of number of participant nucleons $<N_{part}>$. Both ratios indicate an enhancement with respect to PYTHIA and they are well described by coalescence models. This points to possible charm quark redistribution happening in $Au+Au$ collisions, which enhances these ratios. Furthermore, it is more likely to produce a strange open charm hadrons, due to strangeness enhancement~\cite{bib:STAR:strangeness_enhancement}.

\begin{figure}[h]
\centering
	\subfloat{		
		{\includegraphics[width=0.48\textwidth]{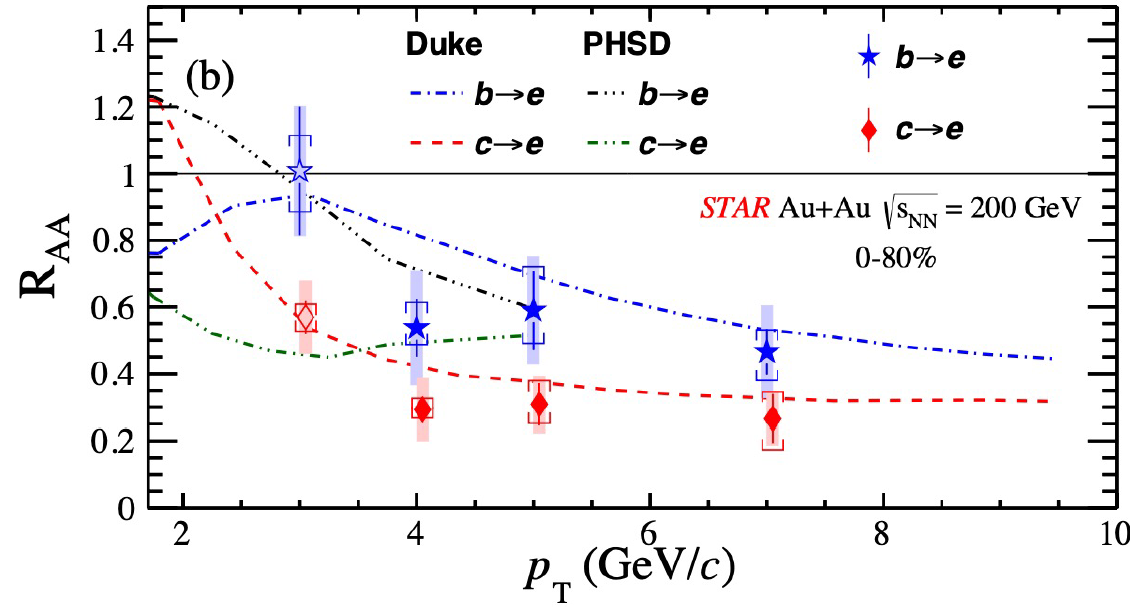}
		\label{fig:cbe:Raa}}
		{\includegraphics[width=0.52\textwidth]{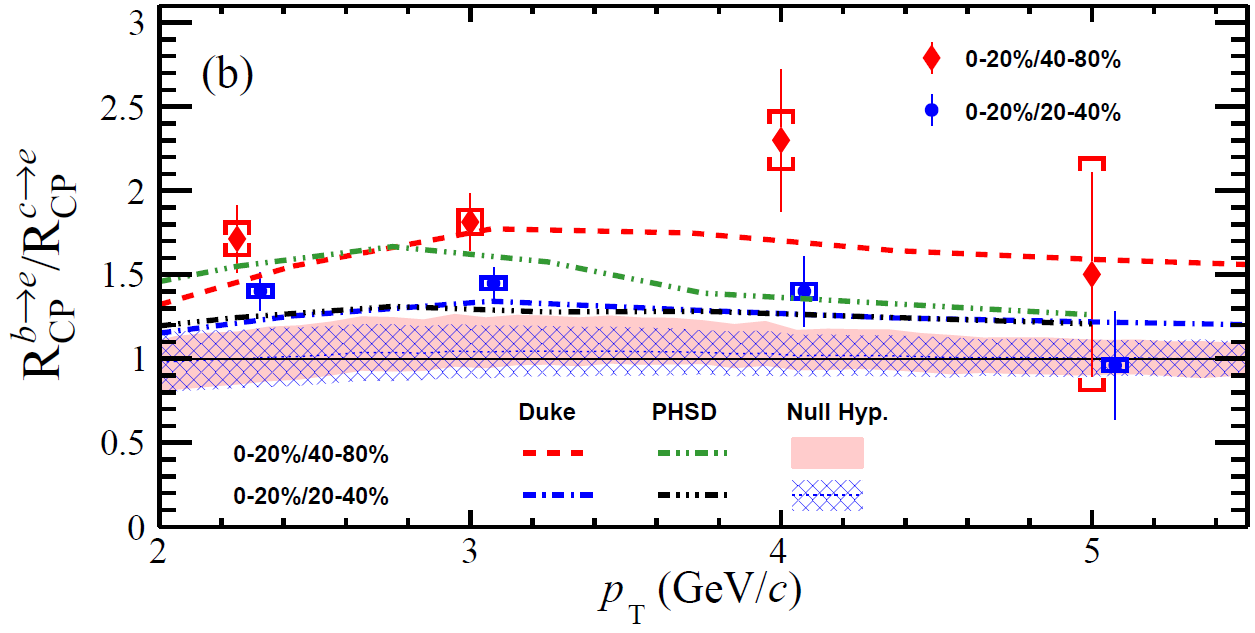}
		\label{fig:cbe:Rcp}}
		}
\caption{Left: Nuclear modification factor of $c \rightarrow e$ and $b \rightarrow e$ decays measured vs. $p_{T}$ by STAR in $A+A$ collisions at $200\:\mathrm{GeV}$~\cite{bib:STAR:cbe}.
Right: Ratio of $R_{CP}$ of bottom to charm semileptonic decays vs. $p_{T}$.~\cite{bib:STAR:cbe}
}
\label{fig:cbe:Raa:all}       % Give a unique label
\end{figure}

\subsection{Charm and bottom energy loss}

The nuclear modification factor $R_{AA}$ of $c \rightarrow e$ and $b \rightarrow e$ decays in $Au+Au$ collisions at $200\:\mathrm{GeV}$ was measured by STAR~\cite{bib:STAR:cbe}. The $p_{T}$ dependence of $R_{AA}$ is shown in Fig.~\ref{fig:cbe:Rcp} (left), while the ratio of $R_{CP}$ of $b$ to $c$ is presented in Fig.~\ref{fig:cbe:Rcp} (right). The data show a mass ordering of heavy quark energy loss, where charm quarks are more suppressed than bottom. This can be related to the observed "`dead cone"' effect~\cite{DeadConeEffect_obs}. A similar measurement was performed at PHENIX~\cite{bib:cbe:PHENIX:AuAu}, yielding consistent results. The PHENIX and STAR data are well described by the energy loss models for $p_{T}>4\mathrm{GeV/c}$, which include both collisional and radiative energy loss.

\begin{figure}[h]
\centering
	\subfloat{		
		{\includegraphics[width=0.5\textwidth]{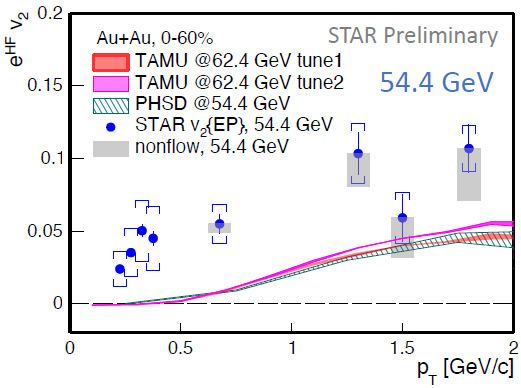}
		\label{fig:npe:STAR:v2:54gev}}
		{\includegraphics[width=0.5\textwidth]{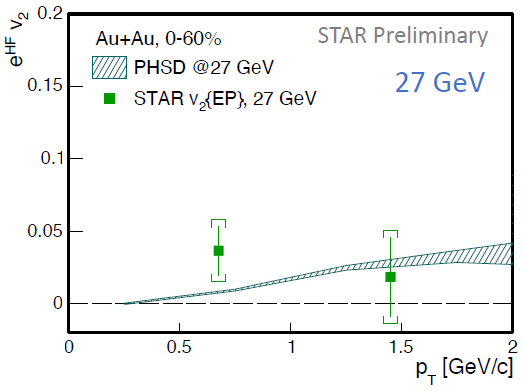}
		\label{fig:npe:STAR:v2:27gev}}
		}
\caption{Left: Elliptic flow $v_{2}$ of heavy flavor decay electrons vs. $p_{T}$ measured by STAR at $54.4\:\mathrm{GeV}$~\cite{bib:npe:STAR:v2:54_27GeV}.
Right: Similar measurement for $27\:\mathrm{GeV}$~\cite{bib:npe:STAR:v2:54_27GeV}.}
\label{fig:npe:STAR:v2}       % Give a unique label
\end{figure}

\subsection{Flow of open heavy flavor}

Measurements of flow of open heavy flavor, especially elliptic flow $v_{2}$ provide insight into heavy quark thermalization and interactions. This was studied at PHENIX~\cite{bib:cbe:PHENIX:ppAuAu} and STAR~\cite{bib:charm:STAR:v2} in $Au+Au$ collisions at $200\:\mathrm{GeV}$. There is a significant $v_{2}$ for $D^{0}$ mesons, which has a similar magnitude to lighter hadrons and follows scaling with number of constituent quarks. This means that charm quarks interact strongly with the medium and that they may be thermalized at RHIC energy. Similar studies have been performed in the semileptonic decay channel~\cite{bib:npe:STAR:v2} at $200, 62.4, 39\:\mathrm{GeV}$ and at $54.4, 27\:\mathrm{GeV}$~\cite{bib:npe:STAR:v2:54_27GeV} by STAR. The measured $v_{2}$ vs. $p_{T}$ is presented in Fig.~\ref{fig:npe:STAR:v2} and compared to model calculations. Strong heavy flavor interactions with medium persist at these energies, however the data for $p_{T}<1.4\:\mathrm{GeV/c}$ are underestimated by the TAMU and PHSD calculations~\cite{bib:TAMU:npe:v2, bib:PHSD:npe:v2, bib:PHSD:npe:v2_2}.

\section{Quarkonium}

\subsection{Quarkonium production and polarization}

Both $J/\psi$~\cite{bib:Jpsi:pp:STAR:mult, bib:Jpsi:pp:STAR:500} and $\varUpsilon$~\cite{bib:Kosarzewski2021} spectra were measured by STAR and $J/\psi$ at PHENIX~\cite{bib:Jpsi:pp:PHENIX} in $p+p$ collisions. The data are overall well described by model calculations~\cite{bib:Frawley2008,bib:Jpsi:NRQCD:NLO,bib:Butenschoen2012,bib:jpsi_cgc,bib:jpsi:NRQCDtevatron,bib:upsCGC,bib:jpsi_cgc} with a contribution from $B\rightarrow J/\psi$ meson decays calculated with FONLL~\cite{bib:FONLL:Cacciari2012,bib:FONLL:Cacciari2015}. The only exception is $\varUpsilon$ $p_{T}$ spectrum, which is overestimated by CGC+NRQCD calculation.

 Another way to study quarkonium production mechanism is to measure it's polarization. The $J/\psi$ polarization was measured by STAR and PHENIX in $p+p$ collisions at $200\:\mathrm{GeV}$~\cite{bib:Jpsi:STAR:Pol200GeV} and $510\:\mathrm{GeV}$~\cite{bib:Jpsi:PHENIX:Pol510GeV} respectively in different reference frames. The results for polarization parameters are consistent with no polarization and the same results are obtained between the reference frames. The only exception is the value of $\lambda_{\theta}$ at high $p_{T}$ and $|y|<0.5$. The uncertainties are large, which means it's hard to rule out models, but the best description of the data is given by CGC+NRQCD calculation. PHENIX was able to measure polarization at forward rapidity $1.2<y<2.2$ and there is no difference compared to $|y|<0.35$.

\subsection{Quarkonium suppression}

\begin{figure}[h]
\centering
		\includegraphics[width=0.5\textwidth]{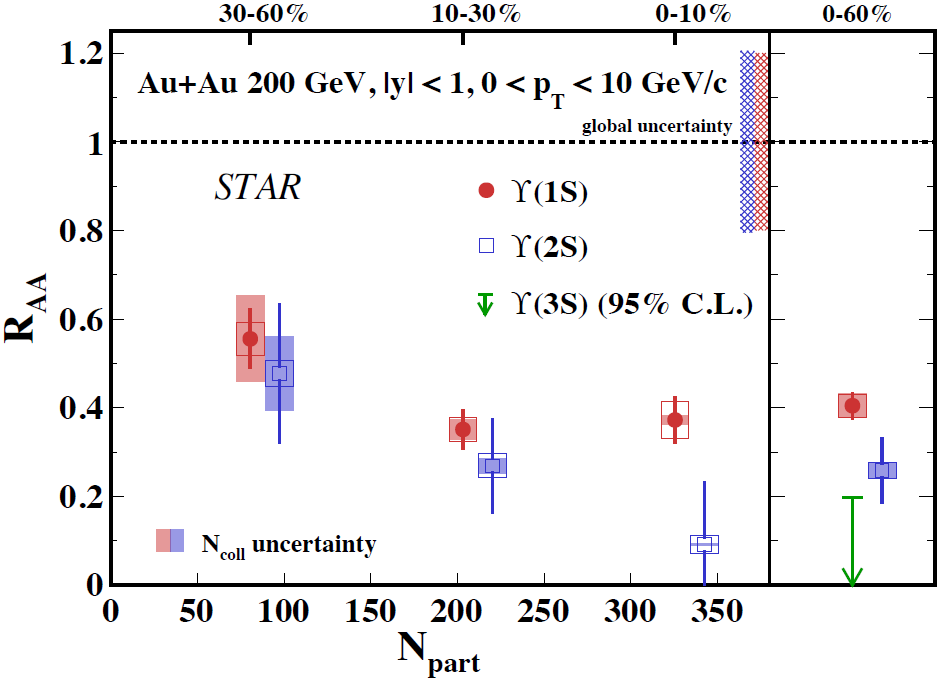}
\caption{Nuclear modification factor of $\varUpsilon$ states vs. $N_{part}$~\cite{bib:Ups:STAR:states}. }
\label{fig:Ups:STAR:Raa:all}       % Give a unique label
\end{figure}

Suppression of $J/\psi$ was measured using nuclear modification factor by STAR~\cite{bib:Jpsi:STAR:AuAu:lowpt} and PHENIX~\cite{bib:JPsi:PHENIX:200GeV} in $Au+Au$ collisions at $200\:\mathrm{GeV}$.
The $R_{AA}$ of $J/\psi$ was also measured at $62.4\:\mathrm{GeV}$ and $39\:\mathrm{GeV}$ by both experiments~\cite{bib:Jpsi:AuAu:STAR:BES, bib:JPsi:PHENIX:62_39GeV}.
The data show a trend of increasing $R_{AA}$ with $p_{T}$ and decreasing with centrality. In addition, no significant energy dependence is observed within uncertainties up to collision energy of $200\:\mathrm{GeV}$. This behavior is reproduced by the suppression model calculations including regeneration of $J/\psi$~\cite{bib:Jpsi:Rapp}.

STAR has also measured $R_{AA}$ vs. $<N_{part}>$ and $v_{2}$ of $J/\psi$ in $Ru+Ru$ and $Zr+Zr$ collisions~\cite{bib:Jpsi:STAR:isobar}. The trend is similar across different systems, which may indicate an interplay of dissociation, regeneration and cold nuclear matter effects. The measured $v_{2}$ is consistent with no flow below $p_{T}<4\:\mathrm{GeV/c}$.

\begin{figure}[h]
\centering
	\subfloat{		
		{\includegraphics[width=0.5\textwidth]{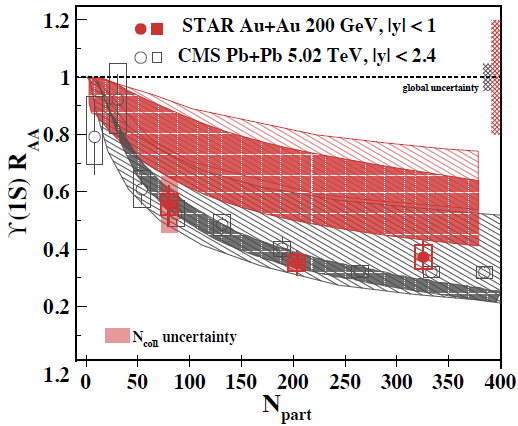}
		\label{fig:Ups:STAR:Raa:models:1S}}
		{\includegraphics[width=0.5\textwidth]{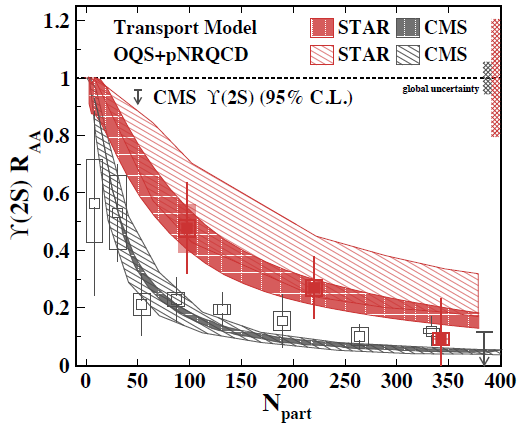}
		\label{fig:Ups:STAR:Raa:models:2S}}
		}
\caption{Nuclear modification factor $R_{AA}$ vs. $N_{part}$ for $\varUpsilon(1S)$ (left) and $\varUpsilon(2S)$ right measured by STAR  and CMS~\cite{bib:Ups:STAR:states}. }
\label{fig:Ups:STAR:Raa:models}       % Give a unique label
\end{figure}

Nuclear modification factor of $\varUpsilon$ states was also measured by STAR~\cite{bib:Ups:STAR:dAu, bib:Ups:STAR:dAuErr, bib:Ups:STAR:states} in $Au+Au$ collisions at $200\:\mathrm{GeV}$. This is shown vs. $N_{part}$ in Fig.~\ref{fig:Ups:STAR:Raa:all}. It is a first observation of sequential suppression of $\varUpsilon$ states at RHIC energy. Results also indicate increasing suppression with centrality. The $R_{AA}$ of $\varUpsilon(1S)$ and $\varUpsilon(2S)$ in Fig.~\ref{fig:Ups:STAR:Raa:models} is also compared to the model calculations, which include dissociation, regeneration and cold nuclear matter effects~\cite{PhysRevC.96.054901,Brambilla_2022}. The models are consistent with the data, but show larger separation between STAR and CMS. The $R_{AA}$ of $\varUpsilon(1S)$ is similar at RHIC and LHC, while $\varUpsilon(2S)$ indicates a smaller suppression.

\section{Summary}

So far, STAR and PHENIX performed a wide range of heavy flavor measurements. The charm $R_{AA}$ indicate a similar level of suppression as for light hadrons and $v_{2}$ is also similar between heavy flavor and light hadrons. This means that there is strong interaction of charm quarks with medium down to $54.4\:\mathrm{GeV}$. The $R_{AA}$ of bottom is larger than charm, which suggests a mass dependence.

Quarkonium production at RHIC is well described by the production models with a few exceptions. The $J/\psi$ $R_{AA}$ has similar value when measured in different colliding systems, so there may be an interplay of suppression, regeneration and cold nuclear matter effects.
Both $J/\psi$ and $\varUpsilon$ $R_{AA}$ exhibit a similar trend of increasing suppression with centrality and decreasing with $p_{T}$. Interestingly, the $\varUpsilon(1S)$ is similarly suppressed at RHIC and LHC, mostly due to dissociation of feed-down to $\varUpsilon(1S)$ state. Furthermore a sequential suppression of $\varUpsilon$ states was observed by STAR.

\section*{Acknowledgments}

This work was supported by grant from The Czech Science Foundation, grant number: GJ20-16256Y

%\section*{References}

%
\bibliography{Bibliography}

\end{document}